\documentclass[preprints,article,accept,pdftex,moreauthors]{mdpi}
\firstpage{1} 
\pubvolume{1}
\issuenum{1}
\articlenumber{0}
\pubyear{2022}
\copyrightyear{2022}
\datereceived{} 
\dateaccepted{} 
\datepublished{} 
\hreflink{https://doi.org/} 

\usepackage{makecell}

\Title{A PeVatron Candidate: Modelling the Boomerang Nebula in X-ray Band}

\TitleCitation{A PeVatron Candidate: Modelling the Boomerang Nebula in X-ray Band}

\Author{Xuan-Han Liang $^{1,2}$\orcidA{}, Chao-Ming Li$^{1,2}$, Qi-Zuo Wu $^{1,2}$, Jia-Shu Pan$^{1,2}$ and Ruo-Yu Liu $^{1,2,}$*\orcidB{}}

\AuthorNames{Xuan-Han Liang, Chao-Ming Li, Qi-Zuo Wu, Jia-Shu Pan and Ruo-Yu Liu}

\AuthorCitation{Liang, X.-H.; Li, C.-M..; Wu, Q.-Z.; Pan, J.-S.; Liu, R.-Y.}

\address{%
$^{1}$ \quad Department of Astronomy, Nanjing University, 163 Xianlin Avenue, Nanjing 210023, People’s Republic of China\\
$^{2}$ \quad Key Laboratory of Modern Astronomy and Astrophysics, Nanjing University, Ministry of Education, Nanjing 210093, People’s Republic of China}

\corres{Correspondence: ryliu@nju.edu.cn}

\abstract{Pulsar wind nebula (PWN) Boomerang and the associated supernova remnant (SNR) G106.3+2.7 are among candidates for the ultra-high-energy (UHE) gamma-ray counterparts published by LHAASO. Although the centroid of the extended source, LHAASO~J2226+6057, deviates from the pulsar's position by about $0.3^\circ$, the source partially covers the PWN. Therefore, we cannot totally exclude the possibility that a part of the UHE emission comes from the PWN. Previous studies mainly focus on whether the SNR is a PeVatron, while neglecting the energetic PWN. Here, we explore the possibility of the Boomerang Nebula being a PeVatron candidate by studying its X-ray radiation. By modelling the diffusion of relativistic electrons injected in the PWN, we fit the radial profiles of the X-ray surface brightness and the photon index. The solution with a magnetic field $B=140\,\upmu\text{G}$ can well reproduce the observed profiles and implies a severe suppression of IC scattering of electrons. Therefore, a proton component need be introduced to account for the UHE emission, in light of recent LHAASO's measurement on Crab Nebula, if future observations reveal part of the UHE emission originating from the PWN. In this sense, Boomerang Nebula would be a hadronic PeVatron.}


\begin{document}

\section{Introduction}

Cosmic rays (CRs) are mostly protons and nuclei moving nearly at the speed of light. Their origins are not totally clear because as charged particles their trajectories are deflected by magnetic fields when travelling from sources to the observer. The "knee" structure of the CR spectrum shows up at about 3\,PeV, and it is generally believed that CRs with energy below the structure are dominated by the Milky Way. Hence, CR protons can be accelerated up to PeV in Galactic sources, and such CR factories are called PeVatrons. Recently, LHAASO has published a list of ultra-high-energy (UHE, $E_\gamma >100\,$TeV) gamma-ray sources, which are considered as Galactic PeVatron candidates, and for the very first time we can study such powerful astrophysical objects in the PeV regime \cite{cao2021}.\par

LHAASO~J2226+6057 is one of the UHE gamma-ray sources. SNR G106.3+2.7, firstly discovered by a radio survey of the Galactic plane \cite{joncas1990}, is a possible counterpart. Later this elongated Galactic source was interpreted as a supernova remnant (SNR) with two distinct components, the head and the tail \cite{pineault2000}. The relatively compact head region of high surface brightness is located in the northeast of the comet-shaped structure, while the diffuse tail region of low surface brightness extends to the southwest. From a bigger picture, the tail region is believed to break into a large low-density HI bubble, which may be created by the stellar wind and/or previous supernova explosions \cite{kothes2001}. The most prominent structure of the head region is the pulsar wind nebula (PWN) Boomerang in the north, which is powered by the young and powerful pulsar PSR J2229+6114 with a characteristic age of 10 kyr and a high spin-down luminosity of $2.2\times10^{37}$ erg s$^{-1}$ \cite{halpern2001a}. A distance of 3 kpc of the system is estimated based on the X-ray absorption \cite{halpern2001b}, but a shell-like structure of HI wrapping around the SNR head, which is suggested to be interacting with the PWN-SNR complex, implies a physical distance of 800 pc of the system based on the velocity of the HI emission \cite{kothes2001}.  On the other hand, the molecular clouds found in the complex seem to have a different spatial distribution, with the location coincident with the tail region in our line of sight based on CO observations \cite{kothes2001}. However, it is still uncertain whether the molecular clouds are directly disturbed by the SNR shocks\cite{liu2022}, but they are expected to be nearby and thus can still be illuminated by the protons that escaped from the SNR \cite{ge2021,bao2021}.\par

Because of the special morphology, the PWN-SNR complex has drawn attention and multi-wavelength data have been obtained. Radio observations pioneered, including one especially for the Boomerang Nebula \cite{kothes2006}. The gamma-ray observations cover a large range from GeV to TeV, including the detection by Fermi-LAT in the range of 0.1 - 500 GeV \cite{xin2019,fang2022}, by the Milagro collaboration at 20 TeV \cite{abdo2007} as well as 35 TeV \cite{abdo2009}, and by the VERITAS collaboration in the range of 0.9 - 16 TeV \cite{acciari2009} with source named as VER J2227+608. Recently, HAWC \cite{albert2020} identified the target as HAWC J2227+610 and reported observations of gamma-ray emission up to 100 TeV. The Tibet AS$\gamma$ \cite{amenomori2021} experiments also detected photons up to 100 TeV with measured emission centroid deviating from the pulsar’s location at a confidence level of $3.1\sigma$. Soon LHAASO \cite{cao2021} lifted the maximum photon energy to 500 TeV.\par

The UHE ($E_\gamma\gtrsim$ 100 TeV) gamma-rays are produced via inverse Compton (IC) scattering by electrons/positrons or via $\pi^0$ decay following the proton-proton collision. Several simulations have explored the leptonic or hadronic origin of the UHE emission \cite{liusm2020,bao2021,yu2022,yang2022,breuhaus2022}. Because of the spacial coincidence between the extended gamma-ray emission and the molecular clouds in the tail region, the hadronic origin of the TeV photons is favored (e.g. \cite{amenomori2021,ge2021}). However, as an extended source, LHAASO~J2226+6057 also partially covers the head region. This is also supported by the morphology observed by VERITAS \cite{acciari2009} and a recent preliminary analysis by MAGIC \cite{Magic2022} at relatively lower energies. Therefore, we cannot totally exclude the possibility that the UHE photons, or at least part of them, originate from the PWN. \par

On the other hand, the nonthermal X-ray radiation traces UHE electrons and hence is useful for determining whether the origin of the UHE gamma-ray emission is leptonic or hadronic. Early study has only focused on PSR J2229+6114 and its vicinity \cite{halpern2001b}, but recently, data from Chandra, XMM-Newton and Suzaku have been used to study the whole structure. X-ray spectra from different parts of the region follow power-law distributions \cite{ge2021}, suggesting a thermal synchrotron origin of the X-ray emission. Interpreting the X-ray emission as thermal would invoke an unreasonably low metal abundance \cite{fujita2021}. Ge et al. \cite{ge2021} (hereafter Ge2021) obtained radial profiles of X-ray surface brightness and  photon index as a function of the distance from PSR J2229+6114, with a joint analysis of data from both Chandra and XMM-Newton. Figure 2 in Ge2021 shows the combined images in 1-7 keV with instrumental background subtracted and exposure corrected. Chandra mainly covers PSR J2229+6114, the wind nebula and the nearby head region, while XMM-Newton covers most of the head region and the extended tail region. Figure 3A in Ge2021 shows the two radial profiles. Considering the trend of the profiles, they can be divided into three parts. The boundary between the first and second parts is about $100^{\prime\prime}$ from the pulsar, which corresponds to the radius of the Boomerang Nebula, while the boundary between the second and third parts is about $840^{\prime\prime}$ from the pulsar. In the first part, the surface brightness, or intensity, drops rapidly, while the photon index rises rapidly. This part is thought to be dominated by electrons confined in PWN. In the second part, the intensity continues to decline, but is shallower than that of the first part, and the corresponding index also gently increases. Electrons escaping from the PWN are believed to contribute most in this area, corresponding to the head region. In the third part, the intensity and the photon index no longer change significantly. Due to the rapid cooling of the X-ray emitting electrons, X-ray emission from this part is most likely to be generated by electrons accelerated at the SNR shock. These two profiles show how the X-ray radiation varies spatially with the distance from the pulsar, and their characteristics suggest two different origins of the electrons, namely the PWN and the SNR. 

To further test this hypothesis, a numerical calculation is needed beyond the phenomenological analysis. In this paper, we model the X-ray profiles around the Boomerang Nebula obtained by Ge2021 with diffusing relativistic electron/positron pairs injected in the PWN. The aim of the simulation is to explore the potential of the Boomerang Nebula as a PeVatron candidate.\par
The rest of this paper is organized as follows. The model is introduced in detail in Section \ref{sec:method}. Section \ref{sec:result} presents the fitting result and the spectral energy distribution (SED). We discuss the simulation result and the possible contribution of UHE emission from the Boomerang Nebula in Section \ref{sec:discussion}, and finally draw a conclusion in Section \ref{sec:conclusion}.

\section{Method} \label{sec:method}

\subsection{Two electron components} \label{subsec:m1}

Since electrons have two different origins according to Ge2021, we consider two components in our simulations. For the SNR accelerated electrons, we adopt the fitting results of the tail region from Ge2021. The electron distribution is described by a broken power-law with a super-exponential cutoff
\begin{equation}
    \frac{{\rm d}N_e}{{\rm d}E_e}=N_{0,e}E_e^{-\alpha_e}\left[ 1+\left( \frac{E_e}{E_{e,\rm b}} \right)^\sigma \right]^{-\frac{\Delta\alpha_e}{\sigma}}e^{-\left( \frac{E_e}{E_{e,\rm max}} \right)^2},
\end{equation}
where $\alpha_e$ is the spectral index before the break, $\Delta\alpha_e$ is the index change when the energy exceeds spectral break energy ${E_{e,\rm b}}$, $\sigma$ is the smoothness of the spectral break, and $E_{e, \rm max}$ is the cutoff energy. $N_{0,e}$ is a normalization factor, which is determined by
\begin{equation}
    \int E_e\frac{{\rm d}N_e}{{\rm d}E_e}\,{\rm d}E_e=W_e \label{SNR},
\end{equation}
where the total energy $W_e$ is obtained by fitting the data, and the limit of integration is from 5 MeV to 5 PeV. They use a $B=20\,\upmu\text{G}$ magnetic field and fix $\sigma$ to 5, and other parameters are obtained as $\alpha_e=2.3$, $\Delta\alpha_e=1.4$, ${E_{e,\rm b}}=5\,\text{MeV}$, $E_{e,\rm max}=200\,\text{TeV}$, and $W_e=3\times10^{46}\,\text{erg}$.

For PWN originated electrons, we consider temporal evolution of electrons/positrons (hereafter we do not distinguish positrons from electrons for simplicity) injected from the pulsar's birth (i.e., $t=0$) to the present time (i.e., $t=t_{age}$), following its spin-down history. Assuming the braking index $n=3$, the dissipated rate of pulsar rotational energy, or spin-down luminosity, can be described by
\begin{equation}
    L_t(t)=L_0\left( 1+\frac{t}{\tau_0} \right)^{-2},
\end{equation}
where 
\begin{equation}
    \tau_0\equiv\frac{P_0}{2\dot{P_0}}=\tau_c-t_{age}
\end{equation}
is the initial spin-down time scale of the pulsar. The age of the pulsar is given by
\begin{equation}
    t_{age}=\frac{P}{2\dot{P}}\left[ 1-\left( \frac{P_0}{P} \right)^2 \right],
\end{equation}
where $P_0$ is the initial rotation period of the pulsar. If $P_0\ll P$, $t_{\rm age}$ will reduce to
\begin{equation}
    \tau_c\equiv\frac{P}{2\dot{P}},
\end{equation}
which is called characteristic age and is an estimate of the age of the system. We  assume that the injected electron distribution takes the form of an exponential cutoff power-law, and electrons are injected at a rate of
\begin{equation}
    Q_{\rm inj}\left( E_e,t \right)\equiv\frac{{\rm d}N}{{\rm d}E_e{\rm d}t}=Q_0\left(t\right)E_e^{-p}e^{-\frac{E_e}{E_{\rm cut}}}, \label{injection}
\end{equation}
where $p$ is the spectral index of the injected electron spectrum, and $E_{\rm cut}$ is the cutoff energy. The normalization factor $Q_0\left(t\right)$ is determined by
\begin{equation}
    \int E_e Q_{\rm inj}\left( E_e,t \right)\, {\rm d}E_e=\eta_e  L_t(t),
\end{equation}
where $\eta_e$ is the efficiency at which the energy dissipated by pulsar rotation is converted to electron energy at the termination shock.\par

According to Kothes et al. \cite{kothes2001}, the elongated shape of the source is caused by the special environment. The radial profiles follow the direction of PWN-head-tail, and thus can be assumed not to be affected by the possible reverse shock from the dense clouds in the northeast near the pulsar. Therefore, it is acceptable to consider the spherically symmetric propagation of electrons. In addition, we neglect the proper motion of the pulsar in the simulation for simplicity. The pulsar acts as a point source for isotropic injection of electrons and the center of the sphere.\par

The general solution of the diffusion equation is incompatible with the theory of relativity at $r<3D/c$, which leads to the phenomenon of superluminal propagation\citep{aloisio2009}. Because the rigorous solution in relativistic case has not been obtained so far, it is necessary to modify the diffusion solution in non-relativistic case. One method is to introduce ballistic propagation at small radius. Following the approximation adapted by Recchia et al. \cite{recchia2021}, we use a critical time scale $t_c=3D/c^2$ to distinguish ballistic propagation from diffusion propagation, where $D$ is the diffusion coefficient and $c$ is the speed of light. During a travelling time $t = kt_c$, the ballistic and diffusive distances are $r_{ball}=ct=kct_c$ and $r_{diff} \sim \sqrt{Dt}=\sqrt{kDt_c} \sim \sqrt{k}ct_c$, respectively. For $k<1$, i.e. $t \lesssim t_c$, we will get $r_{ball}<r_{diff}$, suggesting the propagation speed in the diffusive regime exceed the speed of light.\par

In our practical calculation, we adopt the generalized $\text{J\"{u}ttner}$ function from Aloisio et al. \citep{aloisio2009} to consider both propagation mechanisms and connect them smoothly, i.e.,
\begin{equation}
    P_J\left( E_e,r,t \right)=\frac{H\left( ct-r \right)}{4\pi \left( ct \right)^3}\frac{1}{\left[ 1-\left( \frac{r}{ct} \right)^2 \right]^2}\frac{y\left( E_e,t \right)}{K_1\left[ y\left( E_e,t \right) \right]}\exp\left[ -\frac{y\left( E_e,t \right)}{\sqrt{1-\left( \frac{r}{ct} \right)^2}} \right], \label{juttner_func}
\end{equation}
where
\begin{equation}
    y\left( E_e,t \right)=\frac{c^2t^2}{2\lambda\left( E_e,t \right)}. \label{juttner_y}
\end{equation}
$H(x)$ is Heaviside function and $K_1(x)$ is the first order modified Bessel function of the second kind. The existence of $H(x)$ avoids $v=r/t>c$ in the $\text{J\"{u}ttner}$ function. To speed up calculation, we use an approximation (see Appendix \ref{app:juttner}).\par

As electrons propagate outward, they suffer energy loss due to synchrotron radiation and IC radiation. The cooling rate is given by \cite{crab2021}
\begin{equation}
    \dot{E_e}=-\frac{4}{3}\sigma_Tc\left( \frac{E_e}{m_ec^2} \right)^2\left\{ U_B+\sum_iU_{{\rm ph},i}\,/\,\left[ 1+\left( \frac{2.82kT_iE_e}{m_e^2c^4} \right)^{0.6} \right]^\frac{1.9}{0.6} \right\}, \label{loss}
\end{equation}
where $\sigma_T$ is Thomson cross section, $m_e$ is the mass of the electron, $c$ is the speed of light, and $k$ is the Boltzmann constant. $U_B=B^2/8\pi$ is the magnetic field energy density, and $U_{{\rm ph},i}$ and $T_i$ are the radiation field energy density and the corresponding temperature of $i$th component respectively. In addition to the CMB black body spectrum, three types of grey body photon fields in the interstellar medium of the Milky Way are included in the calculation of the energy loss caused by IC scattering. The specific parameters are listed in Table \ref{ph_para}.\par

\begin{table}[H]
    \caption{Parameters of photon fields.}
    \label{ph_para}
    \begin{tabularx}{\textwidth}{CCC}
    \toprule
        photon field & $U_{\rm ph}\,\left(\text{eV}\,\text{cm}^{-3}\right)$ & $T\,(\text{K})$ \\ 
        \midrule
        CMB & 0.26 & 2.73 \\
        FIR & 0.4 & 30 \\
        NIR & 0.2 & 500 \\
        OPT & 0.4 & 5000 \\ 
    \bottomrule
    \end{tabularx}
\end{table}

After considering the pulsar injection, electron propagation and energy loss, the electron distribution at $t_{age}$ can be expressed as
\begin{equation}
    N\left( E_e,r \right)=\int_{0}^{t_{age}} Q_{\rm inj}\left( E_g,t \right)P_J\left( E_e,r,t \right)\frac{{\rm d}E_g}{{\rm d}E_e}\,{\rm d}t, \label{distribution}
\end{equation}
where $E_e$ is the electron energy at the moment, $E_g$ is the initial energy of the electron injected at time $t$. Their relation and ${\rm d}E_g/{\rm d}E_e$ from Equation \ref{distribution} can be obtained via Equation \ref{loss}.

\subsection{MCMC modelling} \label{subsec:m3}

Following the description in Section \ref{subsec:m1}, we have obtained the spatial and spectral distributions of electrons that are injected from PWN at the present time. We have four free parameters: the electron injection spectral index $p$, the magnetic field $B$, and $D_0$ and $\delta$ from the energy-dependent diffusion coefficient
\begin{equation}
    D\left(E\right)=D_0\left( \frac{E_e}{100\,\text{TeV}} \right)^\delta. \label{diff_coeff}
\end{equation}
Other parameters are listed in Table \ref{PWN_para}, where Lorentz factor $\gamma_e=\frac{E_e}{m_ec^2}$. Although the initial rotation period $P_0$ is also a free parameter, it barely influences the result unless it is very close to the current period $P=51.6\,$ms. Therefore we fix it at 40\,ms in our modelling, yielding $t_{\rm age}\approx4200$ years.\par

\begin{table}[H]
    \caption{Fixed parameters of the model.}
    \label{PWN_para}
    \begin{tabularx}{\textwidth}{CCC}
    \toprule
        parameter (unit) & value\\ 
        \midrule
        $n$ & 3 \\
        $P\,(\text{s})$ \cite{halpern2001b} & 0.0516 \\
        $P_0\,(\text{s})$ \cite{halpern2001b} & 0.04 \\
        $\dot{P}\,(\text{s\,s}^{-1})$ \cite{halpern2001b} & $7.827\times10^{-14}$ \\
        $L_s\,(\text{erg\,s}^{-1})$ \cite{halpern2001b} & $2.2\times10^{37}$ \\
        $d\,(\text{pc})$ \cite{kothes2001} & 800 \\
        $\gamma_{inj,min}$ & $2\times10^3$ \\
        $\gamma_{cut}$ & $2\times10^9$ \\
        \bottomrule
    \end{tabularx}
\end{table}

After obtaining the electron distribution, we can calculate the X-ray emission at 1-7 keV which is dominated by synchrotron radiation, employing an acceleration method \cite{fouka2013}. The next step is to integrate the emission along the line of sight. It is worth noting that the distribution of relativistic electrons is no longer angular symmetric during the transition from ballistic to diffuse regime due to the introduction of ballistic propagation. We adopt the distribution from Prosekin et al. \cite{prosekin2015} as $\frac{1}{2\pi}M\left(\mu\right)=\frac{1}{2\pi Z\left(x\right)}e^{-\frac{3\left( 1-\cos\mu \right)}{x}}$, where $Z\left(x\right)=\frac{x}{3}\left( 1-e^{-\frac{6}{x}} \right)$, $x\left(E\right)=\frac{rc}{D\left(E\right)}$,
$\cos\mu=-\frac{r^2+s^2-d^2}{2rs}$, $r=\sqrt{s^2+d^2-2sd\cos\theta}$.
$d$ is the distance from the pulsar listed in Table \ref{PWN_para}, $s$ is the distance from the integral position to the observer, $r$ is the radial distance of the integral position from the pulsar, $\theta$ is the angle between the pulsar direction and the line of sight, and $\mu$ is the angle between radial direction and the line of sight. It is obvious that the distribution is energy-dependent. To fit the SBP, we need to further integrate the intensity $I$ over the range of 1-7 keV. For the SNR accelerated electrons described in Section \ref{subsec:m1}, Equation \ref{SNR} yields a uniform electron distribution in the range of 600 arcmin$^2$. Similarly, we can obtain their synchrotron emission and the total intensity at 1-7 keV.\par

When fitting the observed data of radial profiles, we need to consider two kinds of electrons according to the analysis in Section \ref{subsec:m1}. Following the picture raised by Ge2021, the PWN and the SNR dominate the X-ray emission at the head region and the tail region respectively. Thus, we introduce a truncated angular distance $\theta_c$ for the SNR contribution to highlight the difference between the dominant positions of the two kinds of electrons, with the intensity of the SNR multiplied by a factor $e^{-\theta_c/\theta}$. Finally, we have six free parameters in the model: $p$, $B$, $D_0$, $\delta$, $\eta_e$ and $\theta_c$.\par

Apart from the statistic error, we include an additional systematic error of 10\% of the measured value due to the calibration uncertainties of instruments \cite{Schellenberger2015} in the fitting. We use Markov chain Monte Carlo method (MCMC) to find the best fitting parameters with a Python package {\emph{emcee}} \cite{foreman2013}. The range of the six free parameters in the modelling are listed in Table \ref{MCMC_para}, and $B$, $D_0$, $\eta_e$, and $\theta_c$ take logarithmic values based on 10 in MCMC. We have 12 Markov chains in parallel, each running 5000 steps, of which the first 100 steps are regarded as burn-in and not included in the final sample. 

\begin{table}
    \caption{First row: model parameters; second row: searching range of parameters in MCMC; third row: best-fit values and $1\sigma$ uncertainties. Among the six parameters, $B$, $D_0$, $\eta_e$ and $\theta_c$ take the logarithm in MCMC.}
    \label{MCMC_para}
    \begin{tabularx}{\textwidth}{CCCCCCC}
    \toprule
        parameter & $p$ & $B\,\left(\upmu\text{G}\right)$ & $D_0\,\left(\text{cm}^2\,\text{s}^{-1}\right)$ & $\delta$ & $\eta_e$ & $\theta_c\,\left(\text{arcsec}\right)$\\
        \midrule
        range & [1\,,\,3] & [1\,,\,$10^3$] & [$10^{25}$\,,\,$10^{30}$] & [0\,,\,1] & [$10^{-6}$\,,\,1] & [10\,,\,$10^3$] \\
        \midrule
        percentile & $2.1_{-0.18}^{+0.13}$ & $142_{-22}^{+22}$ & \makecell[c]{$1.5_{-0.68}^{+1.1}\times 10^{29}$} & $0.56_{-0.36}^{+0.30}$ & \makecell[c]{$3.8_{-2.1}^{+5.2}\times 10^{-4}$} & $75_{-50}^{+99}$ \\
    \bottomrule
    \end{tabularx}
\end{table}

\section{Result} \label{sec:result}

\subsection{Fitting result of the radial profiles} \label{subsec:r1}

\begin{figure}[H]
\includegraphics[width=\textwidth]{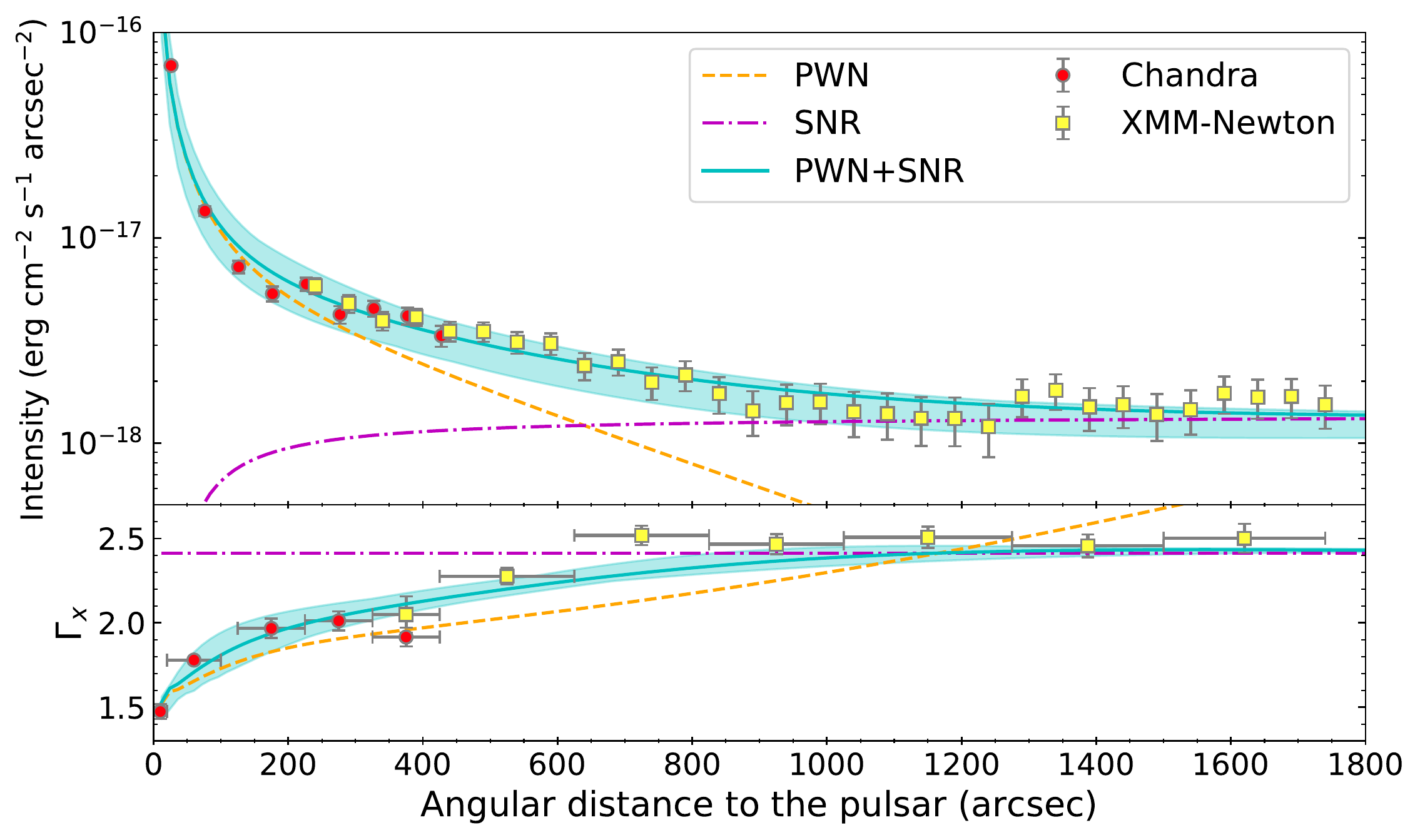}
\caption{The fitting result of the radial profiles of X-ray surface brightness and photon index. The orange dashed line represents the PWN electron component, the purple dash-dotted line represents the SNR electron component, and the cyan solid line with $1\sigma$ error band is the sum of the two. Red and yellow data points are from Chandra and XMM-Newton, respectively, taken from Ge2021. Only Statistic errors are shown.}
\label{chi2}
\end{figure}

We put the medians of the parameter sample from MCMC into the model, and obtain the fitting radial profiles of X-ray surface brightness and photon index of the PWN-SNR complex at 1-7 keV. The y-axis of the upper panel in Figure \ref{chi2} is intensity and that of the lower one is photon index, resembling Figure 3A from Ge2021. The quantiles of each parameter obtained by the MCMC modelling are listed in Table \ref{MCMC_para}. The $\chi^2$ of the fitting is about 41 with 52 degrees of freedom. In this case, the cumulative probability reaches 68.3\% as $1\sigma$ level of the Gaussian distribution when $\Delta \chi^2\approx56$. We fix five parameters, decrease or increase the value of the remaining one, and substitute it into the model so that $\chi^2$ reaches approximately 97. After performing the above procedure on all parameters, we obtain the fitting curves corresponding to 12 groups of parameters. The boundary values enclosed by all curves are taken as the final error band.\par

The fitting curve in Figure \ref{chi2} can well reproduce the profiles of both X-ray surface brightness and photon index. The relatively strong magnetic field $B=142\,\upmu\text{G}$ is worthy of note, as the magnetic field energy density $U_B$ reaches about $500\,\text{eV}/\text{cm}^{3}$, which is much larger than the radiation field energy density listed in Table \ref{ph_para}. The IC scattering of electrons is severely suppressed under such circumstances, and high-energy gamma-ray photons are thus difficult to be produced by PWN accelerated electrons, consistent with the observation of the head region by Fermi-LAT \cite{liusm2020}. Note that in an earlier study \cite{kothes2006}, a stronger magnetic field of $B=2.6\,$mG for the PWN was proposed through ascribing the spectral break at 4.3\,GHz to the synchrotron cooling.

\subsection{Spectral energy distribution} \label{subsec:r2}

LHAASO detected significant signals of UHE photons well above 100\,TeV in this PWN-SNR complex, and the angular extension of the UHE emission covers part of the head region. Therefore, we cannot completely exclude the possibility of part of the UHE emission originating from Boomerang. The large magnetic field obtained from the fitting of X-ray profiles implies a significant suppression of the inverse Compton radiation of electrons. On the other hand, the LHAASO Collaboration \cite{crab2021} perceived a possible hardening of energy spectrum near 1\,PeV in the Crab Nebula. The standard one-component model could not explain this feature, and an alternative two-component scenarios were proposed, where the additional component could be either electron or proton. Enlightened by their work, we envisage that a second spectral component may also exist in Boomerang and could possibly account for part of the UHE emission detected by LHAASO in this region. It may be worth mentioning that although the leptonic origin of gamma rays is the standard picture for PWN, the hadronic origin has been suggested and studied by previous studies \cite[e.g.][]{Atoyan1996,Amato2003, Zhang09, Li2010, DiPalma2017, Liu2021} \par

To check the possibility of this scenario, we assume two particle components in the PWN. The first electron component, which account for the X-ray emission, follows a power-law distribution with exponential cutoff, i.e.,
\begin{equation}
    \frac{{\rm d}N_e}{{\rm d}E_e}=N_{0,\,e}E_{e}^{-\alpha_e}e^{-\frac{E_e}{E_{e,\rm \,cut}}}. \label{PWN_e}
\end{equation}
Note that this component is supposed to be the same as the one modeled in Section 3.1 for the X-ray emission, after spatially integration. Similar to LHAASO's treatment, we also consider an additional electron component (i.e., the two-component leptonic scenario) or a proton component (i.e., the lepto-hadronic scenario) as the second particle component. The emission of the second component is set to contribute half of the measured flux at 500\,TeV by LHAASO. 

\begin{figure}[H]
\begin{adjustwidth}{-\extralength}{0cm}
\centering
\includegraphics[width=18cm]{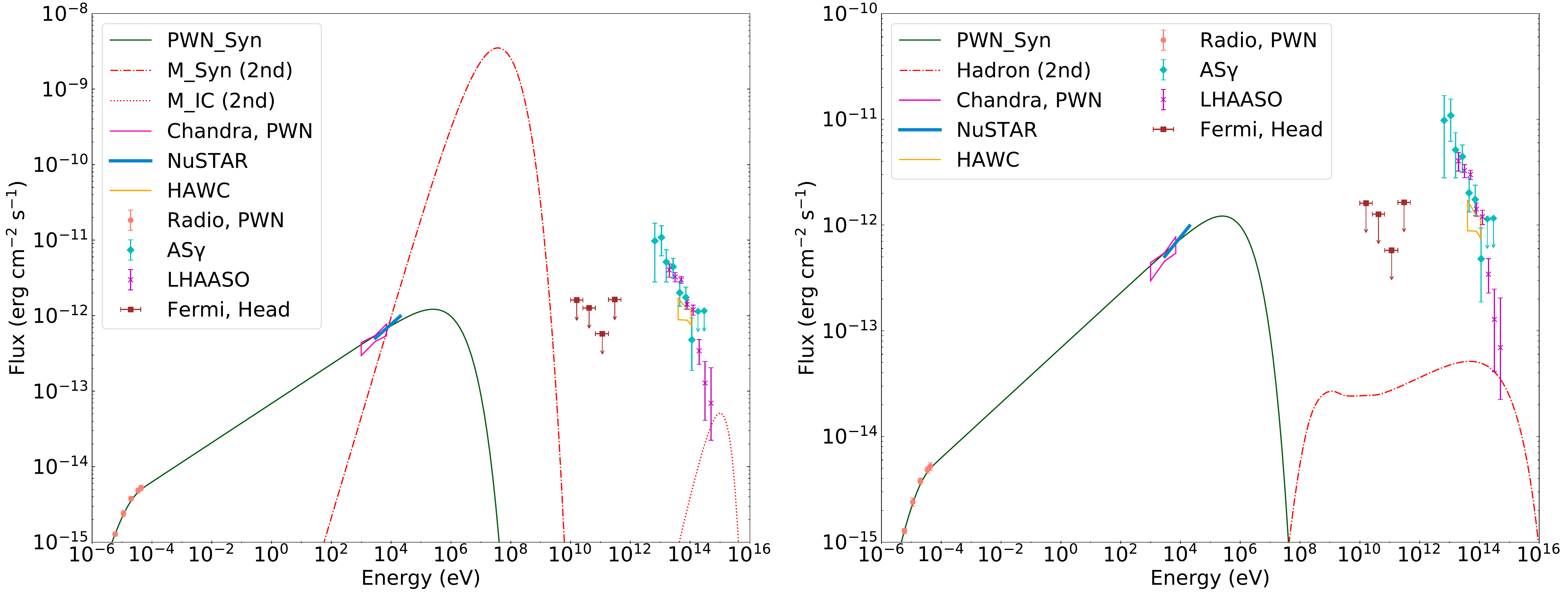}
\end{adjustwidth}
\caption{Left panel: the two-component leptonic scenario for the head region. The dark green solid line is the synchrotron radiation of the power-law type electrons, and the red dash-dotted and dotted lines are the synchrotron and IC radiation of the Maxwellian type electrons respectively. The salmon points on the left are the radio data from Kothes et al. \cite{kothes2006}, and the bright pink butterfly plot is from the observation of Chandra in Ge2021. The recent NuSTAR measurement at 3-20 keV is also shown in cerulean \cite{Nustar2022}. The brown upper limit of the GeV band comes from Fermi-LAT observations of the head region \cite{liusm2020}, while the cyan and magenta TeV data points and orange butterfly plot on the right are from AS$\gamma$ \cite{amenomori2021}, LHAASO \cite{cao2021} and HAWC \cite{albert2020}, respectively. The GeV -- TeV observations are not involved in the fitting since most of them are spatially coincident with the tail region, and are shown for reference only. Right panel: the lepto-hadronic scenario. Dark green solid line represents the synchrotron radiation of electrons, red dash-dot line represents gamma-ray radiation from $\pi^0$ decay.}
\label{2comp}
\end{figure}

In the former scenario, the second electron component is assumed to follow the Maxwellian distribution written as
\begin{equation}
    \frac{{\rm d}N_m}{{\rm d}E_m}=N_{0,\,m}E_m^2e^{-\frac{E_m}{E_{m,\,\rm cut}}}. \label{Max}
\end{equation}
The generated energy spectrum is shown in Figure \ref{2comp}, using the previously obtained best-fit magnetic field of $B=142\,\upmu\text{G}$ and photon fields listed in Table \ref{ph_para}. Other fitting parameters are listed in Table \ref{2comp_para}.  The rather weak IC radiation of the Maxwellian type electrons appears near 100 TeV, peaks around 1 PeV, and drops rapidly. Remarkably, the synchrotron radiation of the Maxwellian type forms a significant bump in the MeV band, peaking at about 100\,MeV. The huge protuberance has a much steeper slope comparing to the measurement from both Chandra and NuSTAR, making this scenario unlikely to be realistic.\par
 
\begin{table}[H]
    \caption{Fitting parameters of the two-component leptonic scenario and lepto-hadronic scenario, corresponding to Figure \ref{2comp}. The former comprises of the PWN electron component (first row) plus the Maxwellian type (second row), and the latter comprises of PWN electron component (first row) plus proton component (third row).}
    \label{2comp_para}
    \begin{tabularx}{\textwidth}{CCCCC}
    \toprule
        {1st $e$} & $\alpha_e$ & $E_{e,\rm \,cut}\,(\text{eV})$ & $E_{e,\rm \,min}\,(\text{eV})$ & $W_e\,(\text{erg})$ \\
        (Equation \ref{PWN_e}) & 2.5 & $4.0\times10^{14}$ & $1.0\times10^9$ & $1.3\times10^{43}$ \\
        \midrule
        {2nd $e$} & $E_{m,\rm \,cut}\,(\text{eV})$ & $E_{m,\rm \,min}\,(\text{eV})$ & $W_m\,(\text{erg})$ \\
        (Equation \ref{Max}) & $4.0\times10^{14}$ & $1.0\times10^{12}$ & $7.5\times10^{42}$ \\
        \midrule
        {$p$} & $\alpha_p$ & $E_{p,\rm \,cut}\,(\text{eV})$ & $W_p\,(\text{erg})$ & $n_{\rm gas}\,\left(\text{cm}^{-3}\right)$ \\
        (Equation \ref{proton}) & 2.0 & $1.0\times10^{16}$ & $1.3\times10^{46}$ & 10 \\
    \bottomrule
    \end{tabularx}
\end{table}

As for the lepto-hadronic scenario, the additional proton component follows the same form as the electron given by
\begin{equation}
     \frac{{\rm d}N_p}{{\rm d}E_p}=N_{0,\,p}E_p^{-\alpha_p}e^{-\frac{E_p}{E_{p,\rm \,cut}}}. \label{proton}
\end{equation}
We follow the semi-analytical method developed by Kafexhiu et al. \cite{kafexhiu2014} to calculate the pionic gamma-ray spectra generated by the proton-proton collisions. The magnetic field and photon fields are the same as the previous case, and the remaining parameters are listed in Table \ref{2comp_para}, where $n_{\rm gas}$ is the gas density. As we can see in Figure \ref{2comp}, gamma-ray emission rises to a certain flux in the GeV band, peaks at several hundred TeV and remains high at 1 PeV before dropping. The proton population still contributes a small amount of energy flux up to 10 PeV compared to the rapid drop in the two-electron case. Assuming the moment of inertia is $10^{45}\,\text{g/cm}^2$ and using the initial and present pulsar period in Table \ref{PWN_para}, the total spin-down energy dissipated since its birth is $4.9\times10^{48}\,\text{erg}$ according to $E_{rot}=\frac{1}{2}I\left(\frac{2\pi}{P}\right)^2$. It is greater than the sum of the energies of the electron and proton, so the scenario does not violate the conservation of energy. Therefore, if the future observations reveal that part of the UHE gamma-ray emission arises from the Boomerang Nebula, our analysis will indicate that Boomerang Nebula, probably as well as other energetic PWNe, is proton PeVatrons.\par

\section{Discussion} \label{sec:discussion}

Although the fitting result of the radial profiles is fairly good, there are some limited aspects of the model. First, we assume a homogeneous magnetic field and diffusion coefficient in the model, which is a simplified assumption and may not be the reality. If variations of the magnetic field or the diffusion coefficient are taken into account, a more complete conclusion will be reached. Second, we only consider diffusive transport of particles in the PWN which has been suggested for many other PWNe \citep{Tang2012,Porth2016,Bao2019,Liu2020_1825,Hu2022}. However, other transport mechanism such as advection may also play an important effect on the electron distribution \cite[e.g.]{KC1984,vanEtten2011,Ishizaki2018}. Last, we cannot ignore the complexity of the surrounding environment of the PWN. The isotropic propagation may be a simplification, as the shape of Boomerang is irregular, and the reverse shock generated by the dense clouds in the northeast may also affect the acceleration and propagation of particles.\par

\begin{figure}[H]
\includegraphics[width=\textwidth]{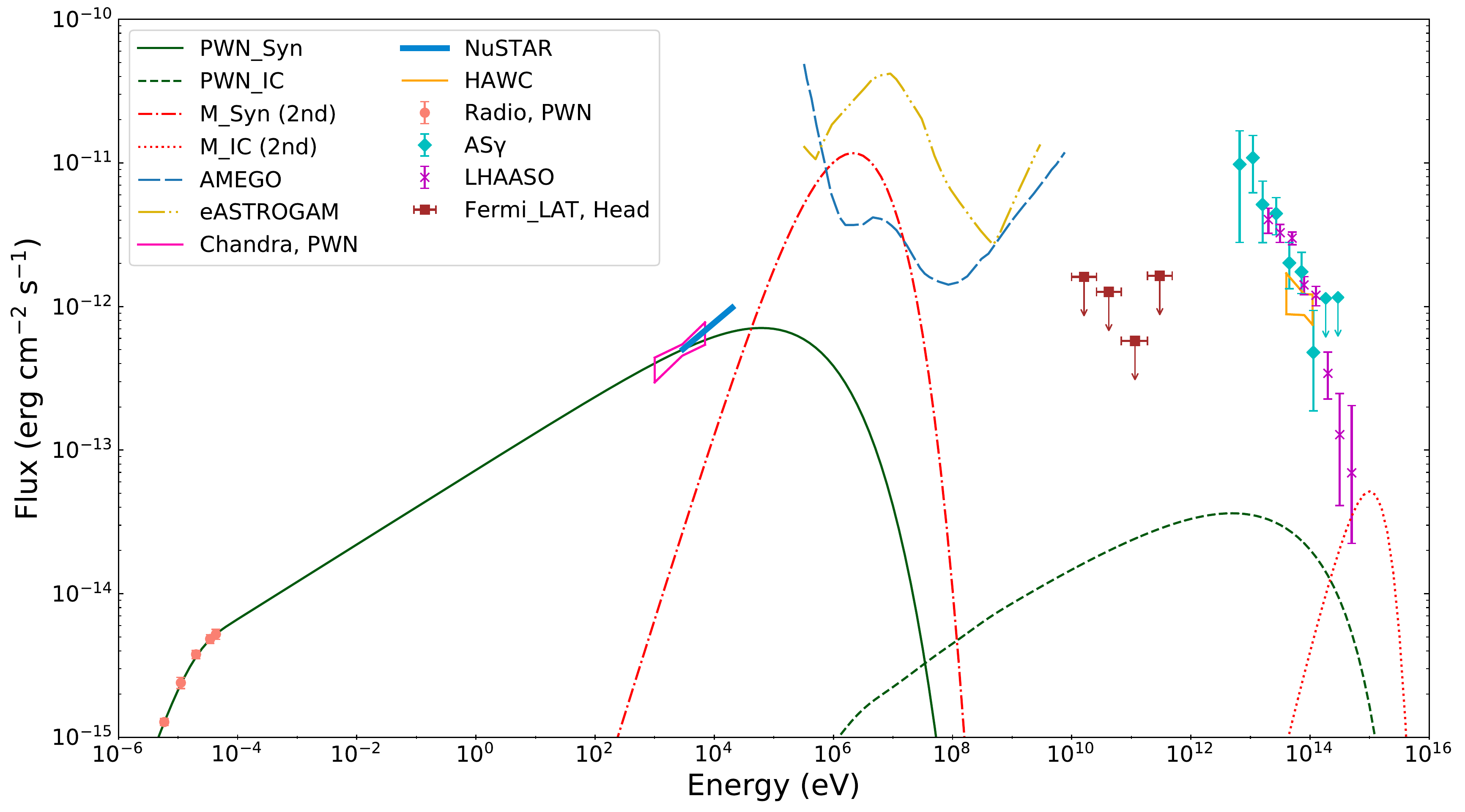}
\caption{The two-component leptonic scenario for the head region. The dark green solid and dashed lines are the synchrotron and IC radiation of the power-law type electrons respectively, and the red dash-dotted and dotted lines are the synchrotron and IC radiation of the Maxwellian type electrons respectively.  The gold dash-dot-doted curve and the blue dashed curve show the 1\,Ms sensitivity of the next-generation MeV gamma-ray detectors e-ASTROGAM \cite{eASTROGAM2017} and AMEGO \cite{AMEGO2017} respectively. Observation data are the same as the ones in Figure \ref{2comp}.}
\label{2e}
\end{figure}

Hence, we try attesting to the result from another perspective. If the electrons from the PWN can substantially contribute to the UHE gamma-ray emission, a weaker magnetic field will be required. Thus, we estimate the lower limit of the magnetic field, with an ideal (and probably unrealistic) assumption that introducing the aforementioned effects in the particle transport can somehow reproduce the profiles of X-ray emissions with a relatively weak magnetic field. In this optimistic limit, a basic requirement for the magnetic field is to make the electrons that emit X-ray photons cool down, so as to explain the rapid increase of the photon index in the PWN region within $100^{\prime\prime}$. Therefore, we have
\begin{equation}
    \tau_c\geq t_{\rm syn}\sim \frac{E_e}{P_{\rm syn}}=\frac{\gamma mc^2}{\frac{4}{3}\sigma_Tc\gamma^2U_B}.
\end{equation}
Here we employ $\tau_c$ instead of $t_{\rm age}$ as a conservative requirement of the cooling timescale in the system. The Lorentz factor can be replaced according to $\nu_c={3\gamma^2eB}/{4\pi mc}$, where $\nu_c$ is the critical frequency of the synchrotron emission. Therefore, to make keV-emitting electrons cool within $\tau_c$, a lower limit of magnetic field $B_{\rm c, LL}=4.7\,\upmu\text{G}$ will be obtained. On the other hand, the Hillas condition \cite{Hillas1984} requires $B\gtrsim E_e/eR$, where $R$ corresponds to the termination shock radius of about 0.1\,pc with a reference to the Crab \cite{gaensler2006,crab2021}. The emitted gamma-ray photon energy via the IC mechanism can be estimated via $E_{IC}\sim E_e\Gamma/(1+\Gamma)$, where $\Gamma={\gamma\epsilon}/{mc^2}$. $\epsilon$ is the average photon energy of the radiation field, and for a black body or grey body radiation it can be expressed as $\epsilon=2.82kT$. Here we use the CMB field as the radiation field. To scatter CMB photons up to $E_{\rm IC}=500$ TeV, the corresponding electron energy need be $E_e\approx 760$ TeV. If we substitute this energy into the Hillas condition, we will get a lower limit of magnetic field $B_{\rm H, LL}=8.2\,\upmu\text{G}>B_{\rm c, LL}$. Similar to the two-component leptonic scenario described in Section \ref{subsec:r2}, we fit the SED again using the $B_{\rm H, LL}$, shown in Figure \ref{2e}. 

We see that with the weaker magnetic field, the IC flux from the first electron component can contribute a small amount of energy flux at several hundred TeV, but it is still far less than the measured values. If we still attribute half of the measured flux at 500 TeV to the second electron component, we still expect a prominent synchrotron bump peaking at MeV regime, although it is not as towering as the one in the left panel of Figure \ref{2comp}. It is not in contrast with the current observation, but the next-generation MeV gamma-ray instruments such as e-ASTROGAM \cite{eASTROGAM2017} and AMEGO \cite{AMEGO2017} will be helpful to discern this scenario. 

\begin{table}[H]
    \caption{Fitting parameters of the two-component leptonic scenario with relatively weaker magnetic field, corresponding to Figure \ref{2e}.}
    \label{2e_para}
    \begin{tabularx}{\textwidth}{CCCCCCC}
    \toprule
        $\alpha_1$ & $E_{1,\,\rm cut}\,(\text{eV})$ & $E_{1,\, \rm min}\,(\text{eV})$ & $W_1\,(\text{erg})$ & $E_{2,\,\rm cut}\,(\text{eV})$ & $E_{2,\,\rm min}\,(\text{eV})$ & $W_2\,(\text{erg})$ \\
        \midrule
        2.5 & $7.6\times10^{14}$ & $4.4\times10^9$ & $9.6\times10^{44}$ & $4.0\times10^{14}$ & $1.0\times10^{12}$ & $7.5\times10^{42}$\\
    \bottomrule
    \end{tabularx}
\end{table}

\section{Conclusion} \label{sec:conclusion}

Based on the radial profiles of X-ray surface brightness and photon index of the PWN-SNR complex at 1-7 keV obtained by Ge2021, we study the X-ray emission of the PWN Boomerang by numerical simulation. In the diffusion framework, we introduce ballistic propagation and use the $\text{J\"{u}ttner}$ function to describe electron transport. Considering the pulsar injection history, the electron transport and the radiation energy loss, we simulate the evolution from the birth of PSR J2229+6114 to the current time of $t_{age}$ to obtain the distribution of PWN accelerated electrons at present. Including both the PWN electrons and the SNR electrons described in Ge2021, we use the MCMC method to fit the radial profiles. The fitting curve obtained by taking the medians of the parameter distribution as the model parameters can reproduce the two profiles at the same time. The median of PWN magnetic field is $B=142\,\upmu\text{G}$. Under the condition of a strong magnetic field, the IC scattering of electrons is severely suppressed and it is difficult to produce gamma-ray radiation, which is consistent with Fermi-LAT observation of the head region of the SNR. However, due to the lack of UHE gamma-ray observations of Boomerang and its vicinity, we cannot rule out the possibility of Boomerang as a PeVatron. \par

Referring to the interpretation by the LHAASO collaboration of the possible hardening of the energy spectrum of the Crab Nebula at 1 PeV, we attempt to simulate the spectra of Boomerang producing UHE gamma-ray radiation by adding a second component of particles, either electrons or protons. For the two-component leptonic scenario, the rather weak IC radiation of the Maxwellian-type electron reaches its peak at 1 PeV and then decreases rapidly. In addition, there is a prominent bump in the MeV band, making this scenario unlikely as it contradicts with the measurement from Chandra and NuSTAR. In the case of adding a proton population, it can account for UHE gamma-ray emission, at least a considerable fraction, with reasonable parameters. The energy spectrum of the proton component extends beyond 1 PeV. In other words, if the a fraction of the UHE gamma-ray photons of LHAASO~J2226+6057 truly comes from Boomerang Nebula, it would indicate that the PWN to be a hadronic PeVatron. This scenario would require detectors like LHAASO to confirm in the future.\par

Furthermore, we discuss the possible scenario with a relatively weaker, albeit unjustified, magnetic field of the PWN. Employing the lower limit of $B=8.2\,\upmu\text{G}$ derived from Hillas condition for the acceleration of electrons emitting 500\,TeV photons. A second electron component of Maxwellian type spectral distribution can interpret the UHE emission via the IC process without violating the current X-ray observations, but it still results in a bump in the spectrum at the MeV band. Next-generation MeV gamma-ray instruments can be helpful to discern this case.

\vspace{6pt} 




\funding{This work is supported by NSFC under grant No.~U2031105.}



\acknowledgments{We thank Chong Ge and Siming Liu for valuable discussions.}

\conflictsofinterest{The authors declare no conflict of interest.} 





\appendixtitles{yes} 
\appendixstart
\appendix
\section{The $\text{J\"{u}ttner}$ function} \label{app:juttner}

The ultrarelativistic particles travel with $v \approx c$ in the ballistic regime. As time goes by, they experience multiple deflections caused by turbulent magnetic field, resulting in the transition to diffusion and the isotropization of the particle directions. We employ the generalized $\text{J\"{u}ttner}$ function to describe this transition, and its original form in Aloisio et al. \cite{aloisio2009} is already given by Equation \ref{juttner_func} and \ref{juttner_y}. The approximation utilized for speeding up is written as
\begin{equation}
    Z\left( y \right)e^y\sim\frac{y}{K_1\left( y \right)}, \label{approximation}
\end{equation}
where
\begin{equation}
    Z\left( y \right)=\frac{y^2}{\left[ 1+\left( \frac{\pi}{2}y \right)^\frac{0.5}{0.53} \right]^{0.53}}.
\end{equation}\par

\begin{figure}
\includegraphics[width=\textwidth]{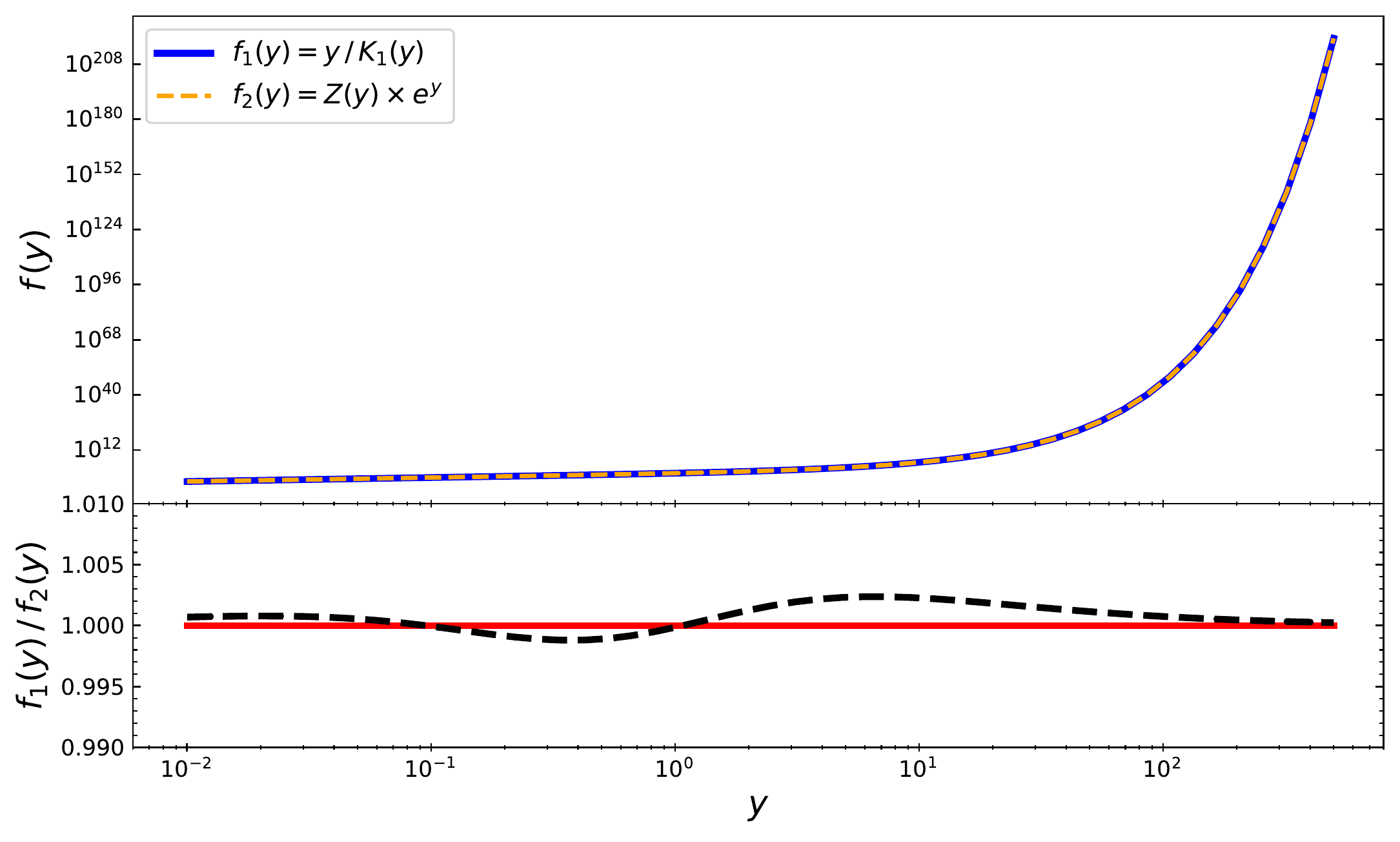}
\caption{Comparison between functions from both sides of Equation \ref{approximation}. The upper panel shows the two functions over five orders of magnitude, with the orange dashed line denoting the left side of Equation \ref{approximation} and the blue solid line denoting the right side of Equation \ref{approximation} from $\text{J\"{u}ttner}$ distribution. The lower panel shows the ratio of the value of the original function to the approximate value of the substitute in black dashed line, with the red solid line being unity for reference.}
\label{2func}
\end{figure}

To verify the accuracy of the approximation, we plot the functions from both sides of Equation \ref{approximation} in Figure \ref{2func}. It is obvious that the values of the two functions are very close over a range of five orders of magnitude. To further demonstrate, the ratio of the two values always fluctuates around unity, with the maximum relative error being only $0.23\%$. Hence, the approximation is accurate enough to be used for speeding up calculation. Eventually the $\text{J\"{u}ttner}$ function we used is written as
\begin{equation}
     P_J\left( E_e,r,t \right)=\frac{H\left( ct-r \right)}{4\pi \left( ct \right)^3}\frac{1}{\left[ 1-\left( \frac{r}{ct} \right)^2 \right]^2}Z\left[ y\left( E_e,t \right) \right]\exp\left[ y\left( E_e,t \right)-\frac{y\left( E_e,t \right)}{\sqrt{1-\left( \frac{r}{ct} \right)^2}} \right].
\end{equation}

\begin{adjustwidth}{-\extralength}{0cm}

\reftitle{References}



\bibliography{template}

\begin{thebibliography}{999}

\bibitem[Cao \em{et~al.}(2021)Cao, Aharonian, An, Bai, Bai, Bao, Bastieri, Bi,
  Bi, Cai, et~al.]{cao2021}
Cao, Z.; Aharonian, F.; An, Q.; Bai, L.; Bai, Y.; Bao, Y.; Bastieri, D.; Bi,
  X.; Bi, Y.; Cai, H.;  et~al.
\newblock Ultrahigh-energy photons up to 1.4 petaelectronvolts from 12
  $\gamma$-ray Galactic sources.
\newblock {\em Nature} {\bf 2021}, {\em 594},~33--36.

\bibitem[Joncas and Higgs(1990)]{joncas1990}
Joncas, G.; Higgs, L.
\newblock The DRAO galactic-plane survey. II-Field at L= 105 deg.
\newblock {\em Astronomy and Astrophysics Supplement Series} {\bf 1990}, {\em
  82},~113--144.

\bibitem[Pineault and Joncas(2000)]{pineault2000}
Pineault, S.; Joncas, G.
\newblock G106. 3+ 2.7: a supernova remnant in a late stage of evolution.
\newblock {\em The Astronomical Journal} {\bf 2000}, {\em 120},~3218.

\bibitem[Kothes \em{et~al.}(2001)Kothes, Uyaniker, and Pineault]{kothes2001}
Kothes, R.; Uyaniker, B.; Pineault, S.
\newblock The supernova remnant G106. 3+ 2.7 and its pulsar-wind nebula: relics
  of triggered star formation in a complex environment.
\newblock {\em The Astrophysical Journal} {\bf 2001}, {\em 560},~236.

\bibitem[Halpern \em{et~al.}(2001{\natexlab{a}})Halpern, Gotthelf, Leighly, and
  Helfand]{halpern2001a}
Halpern, J.P.; Gotthelf, E.; Leighly, K.; Helfand, D.
\newblock A possible X-ray and radio counterpart of the high-energy gamma-ray
  source 3EG J2227+ 6122.
\newblock {\em The Astrophysical Journal} {\bf 2001}, {\em 547},~323.

\bibitem[Halpern \em{et~al.}(2001{\natexlab{b}})Halpern, Camilo, Gotthelf,
  Helfand, Kramer, Lyne, Leighly, and Eracleous]{halpern2001b}
Halpern, J.; Camilo, F.; Gotthelf, E.; Helfand, D.; Kramer, M.; Lyne, A.;
  Leighly, K.; Eracleous, M.
\newblock PSR J2229+ 6114: discovery of an energetic young pulsar in the error
  box of the EGRET source 3EG J2227+ 6122.
\newblock {\em The Astrophysical Journal} {\bf 2001}, {\em 552},~L125.

\bibitem[Liu \em{et~al.}(2022)Liu, Zhou, and Chen]{liu2022}
Liu, Q.C.; Zhou, P.; Chen, Y.
\newblock IRAM 30 m CO-line Observation toward the PeVatron Candidate G106. 3+
  2.7: Direct Interaction between the Shock and the Molecular Cloud Remains
  Uncertain.
\newblock {\em The Astrophysical Journal} {\bf 2022}, {\em 926},~124.

\bibitem[Ge \em{et~al.}(2021)Ge, Liu, Niu, Chen, and Wang]{ge2021}
Ge, C.; Liu, R.Y.; Niu, S.; Chen, Y.; Wang, X.Y.
\newblock Revealing a peculiar supernova remnant G106. 3+ 2.7 as a
  petaelectronvolt proton accelerator with X-ray observations.
\newblock {\em The Innovation} {\bf 2021}, {\em 2},~100118.

\bibitem[Bao and Chen(2021)]{bao2021}
Bao, Y.; Chen, Y.
\newblock On the Hard Gamma-Ray Spectrum of the Potential PeVatron Supernova
  Remnant G106. 3+ 2.7.
\newblock {\em The Astrophysical Journal} {\bf 2021}, {\em 919},~32.

\bibitem[Kothes \em{et~al.}(2006)Kothes, Reich, and Uyan{\i}ker]{kothes2006}
Kothes, R.; Reich, W.; Uyan{\i}ker, B.
\newblock The Boomerang PWN G106. 6+ 2.9 and the magnetic field structure in
  pulsar wind nebulae.
\newblock {\em The Astrophysical Journal} {\bf 2006}, {\em 638},~225.

\bibitem[Xin \em{et~al.}(2019)Xin, Zeng, Liu, Fan, and Wei]{xin2019}
Xin, Y.; Zeng, H.; Liu, S.; Fan, Y.; Wei, D.
\newblock VER J2227+ 608: a hadronic pevatron pulsar wind nebula?
\newblock {\em The Astrophysical Journal} {\bf 2019}, {\em 885},~162.

\bibitem[Fang \em{et~al.}(2022)Fang, Kerr, Blandford, Fleischhack, and
  Charles]{fang2022}
Fang, K.; Kerr, M.; Blandford, R.; Fleischhack, H.; Charles, E.
\newblock Evidence for PeV Proton Acceleration from Fermi-LAT Observations of
  SNR G 106.3+ 2.7.
\newblock {\em Physical Review Letters} {\bf 2022}, {\em 129},~071101.

\bibitem[Abdo \em{et~al.}(2007)Abdo, Allen, Berley, Casanova, Chen, Coyne,
  Dingus, Ellsworth, Fleysher, Fleysher, et~al.]{abdo2007}
Abdo, A.A.; Allen, B.; Berley, D.; Casanova, S.; Chen, C.; Coyne, D.; Dingus,
  B.; Ellsworth, R.; Fleysher, L.; Fleysher, R.;  et~al.
\newblock TeV gamma-ray sources from a survey of the galactic plane with
  Milagro.
\newblock {\em The Astrophysical Journal} {\bf 2007}, {\em 664},~L91.

\bibitem[Abdo \em{et~al.}(2009)Abdo, Allen, Aune, Berley, Chen, Christopher,
  DeYoung, Dingus, Ellsworth, Gonzalez, et~al.]{abdo2009}
Abdo, A.; Allen, B.; Aune, T.; Berley, D.; Chen, C.; Christopher, G.; DeYoung,
  T.; Dingus, B.; Ellsworth, R.; Gonzalez, M.;  et~al.
\newblock Milagro observations of multi-TeV emission from Galactic sources in
  the Fermi bright source list.
\newblock {\em The Astrophysical Journal} {\bf 2009}, {\em 700},~L127.

\bibitem[Acciari \em{et~al.}(2009)Acciari, Aliu, Arlen, Aune, Bautista,
  Beilicke, Benbow, Boltuch, Bradbury, Buckley, et~al.]{acciari2009}
Acciari, V.; Aliu, E.; Arlen, T.; Aune, T.; Bautista, M.; Beilicke, M.; Benbow,
  W.; Boltuch, D.; Bradbury, S.; Buckley, J.;  et~al.
\newblock Detection of extended VHE gamma ray emission from G106. 3+ 2.7 with
  veritas.
\newblock {\em The Astrophysical Journal} {\bf 2009}, {\em 703},~L6.

\bibitem[Albert \em{et~al.}(2020)Albert, Alfaro, Alvarez, Camacho,
  Arteaga-Vel{\'a}zquez, Arunbabu, Rojas, Solares, Baghmanyan, Belmont-Moreno,
  et~al.]{albert2020}
Albert, A.; Alfaro, R.; Alvarez, C.; Camacho, J.A.; Arteaga-Vel{\'a}zquez, J.;
  Arunbabu, K.; Rojas, D.A.; Solares, H.A.; Baghmanyan, V.; Belmont-Moreno, E.;
   et~al.
\newblock HAWC J2227+ 610 and its association with G106. 3+ 2.7, a new
  potential galactic PeVatron.
\newblock {\em The Astrophysical Journal Letters} {\bf 2020}, {\em 896},~L29.

\bibitem[Amenomori \em{et~al.}(2021)Amenomori, Bao, Bi, Chen, Chen, Chen, Chen,
  Chen, Cirennima, Danzengluobu, et~al.]{amenomori2021}
Amenomori, M.; Bao, Y.; Bi, X.; Chen, D.; Chen, T.; Chen, W.; Chen, X.; Chen,
  Y.; Cirennima, C.; Danzengluobu, D.;  et~al.
\newblock Potential PeVatron supernova remnant G106. 3+ 2.7 seen in the
  highest-energy gamma rays.
\newblock {\em Nature Astronomy} {\bf 2021}, {\em 5},~460--464.

\bibitem[Liu \em{et~al.}(2020)Liu, Zeng, Xin, and Zhu]{liusm2020}
Liu, S.; Zeng, H.; Xin, Y.; Zhu, H.
\newblock Hadronic versus leptonic models for $\gamma$-ray emission from VER
  J2227+ 608.
\newblock {\em The Astrophysical Journal Letters} {\bf 2020}, {\em 897},~L34.

\bibitem[Yu \em{et~al.}(2022)Yu, Wu, Wen, and Fang]{yu2022}
Yu, H.; Wu, K.; Wen, L.; Fang, J.
\newblock A leptonic model for the $\gamma$-rays coincident with the tail
  region of the supernova remnant G106. 3+ 2.7.
\newblock {\em New Astronomy} {\bf 2022}, {\em 90},~101669.

\bibitem[Yang \em{et~al.}(2022)Yang, Zeng, Bao, and Zhang]{yang2022}
Yang, C.; Zeng, H.; Bao, B.; Zhang, L.
\newblock Possible hadronic origin of TeV photon emission from SNR G106. 3+
  2.7.
\newblock {\em Astronomy \& Astrophysics} {\bf 2022}, {\em 658},~A60.

\bibitem[Breuhaus \em{et~al.}(2022)Breuhaus, Reville, and Hinton]{breuhaus2022}
Breuhaus, M.; Reville, B.; Hinton, J.
\newblock Pulsar wind nebula origin of the LHAASO-detected ultra-high energy
  $\gamma$-ray sources.
\newblock {\em Astronomy \& Astrophysics} {\bf 2022}, {\em 660},~A8.

\bibitem[{The MAGIC Collaboration} \em{et~al.}(2022){The MAGIC Collaboration},
  {Acciari}, {Ansoldi}, {Antonelli}, {Arbet Engels}, {Artero}, {Asano},
  {Baack}, {Babic}, {Baquero}, {Barres de Almeida}, {Barrio}, {Batkovi{\'c}},
  {Becerra Gonzalez}, {Bednarek}, {Bellizzi}, {Bernardini}, {Bernardos},
  {Berti}, {Besenrieder}, {Bhattacharyya}, {Bigongiari}, {Biland}, {Blanch},
  {B{\"o}kenkamp}, {Bonnoli}, {Bosnjak}, {Busetto}, {Carosi}, {Ceribella},
  {Cerruti}, {Chai}, {Chilingarian}, {Cikota}, {Colak}, {Colombo}, {Contreras},
  {Cortina}, {Covino}, {D'Amico}, {D'Elia}, {da Vela}, {Dazzi}, {de Angelis},
  {de Lotto}, {Delfino}, {Delgado}, {Delgado Mendez}, {Depaoli}, {di Pierro},
  {di Venere}, {Do Souto Espi{\~n}eira}, {Dominis Prester}, {Donini}, {Dorner},
  {Doro}, {Elsaesser}, {Fallah Ramazani}, {Fattorini}, {Fonseca}, {Font},
  {Fruck}, {Fukami}, {Fukazawa}, {Garc{\'\i}a L{\'o}pez}, {Garczarczyk},
  {Gasparyan}, {Gaug}, {Giglietto}, {Giordano}, {Gliwny}, {Godinovic}, {Green},
  {Green}, {Hadasch}, {Hahn}, {Heckmann}, {Herrera}, {Hoang}, {Hrupec},
  {H{\"u}tten}, {Inada}, {Ishio}, {Iwamura}, {Jim{\'e}nez Mart{\'\i}nez},
  {Jormanainen}, {Jouvin}, {Karjalainen}, {Kerszberg}, {Kobayashi}, {Kubo},
  {Kushida}, {Lamastra}, {Lelas}, {Leone}, {Lindfors}, {Linhoff}, {Lombardi},
  {Longo}, {Lopez-Coto}, {L{\'o}pez-Moya}, {L{\'o}pez-Oramas}, {Loporchio},
  {Machado de Oliveira Fraga}, {Maggio}, {Majumdar}, {Makariev}, {Mallamaci},
  {Maneva}, {Manganaro}, {Mannheim}, {Maraschi}, {Mariotti}, {Martinez},
  {Mazin}, {Menchiari}, {Mender}, {Mi{\'c}anovi{\'c}}, {Miceli}, {Miener},
  {Miranda}, {Mirzoyan}, {Molina}, {Moralejo}, {Morcuende}, {Moreno},
  {Moretti}, {Nakamori}, {Nava}, {Neustroev}, {Nigro}, {Nilsson}, {Nishijima},
  {Noda}, {Nozaki}, {Ohtani}, {Oka}, {Otero-Santos}, {Paiano}, {Palatiello},
  {Paneque}, {Paoletti}, {Paredes}, {Pavleti{\'c}}, {Pe{\~n}il}, {Persic},
  {Pihet}, {Prada Moroni}, {Prandini}, {Priyadarshi}, {Puljak}, {Rhode},
  {Rib{\'o}}, {Rico}, {Righi}, {Rugliancich}, {Sahakyan}, {Saito}, {Sakurai},
  {Satalecka}, {Saturni}, {Schleicher}, {Schmidt}, {Schweizer}, {Sitarek},
  {{\v{S}}nidari{\'c}}, {Sobczy{\'n}ska}, {Spolon}, {Stamerra},
  {Stri{\v{s}}kovi{\'c}}, {Strom}, {Strzys}, {Suda}, {Suri{\'c}}, {Takahashi},
  {Takeishi}, {Tavecchio}, {Temnikov}, {Terzic}, {Teshima}, {Tosti}, {Truzzi},
  {Tutone}, {Ubach}, {van Scherpenberg}, {Vanzo}, {Vazquez Acosta}, {Ventura},
  {Verguilov}, {Vigorito}, {Vitale}, {Vovk}, {Will}, {Wunderlich}, {Yamamoto},
  and {Zari{\'c}}]{Magic2022}
{The MAGIC Collaboration}.; {Acciari}, V.A.; {Ansoldi}, S.; {Antonelli}, L.A.;
  {Arbet Engels}, A.; {Artero}, M.; {Asano}, K.; {Baack}, D.; {Babic}, A.;
  {Baquero}, A.;  et~al.
\newblock {Resolving the origin of very-high-energy gamma-ray emission from the
  PeVatron candidate SNR G106.3+2.7 using MAGIC telescopes}.
\newblock In Proceedings of the 37th International Cosmic Ray Conference. 12-23
  July 2021. Berlin,  2022, p. 796.

\bibitem[Fujita \em{et~al.}(2021)Fujita, Bamba, Nobukawa, and
  Matsumoto]{fujita2021}
Fujita, Y.; Bamba, A.; Nobukawa, K.K.; Matsumoto, H.
\newblock X-Ray Emission from the PeVatron-candidate Supernova Remnant G106. 3+
  2.7.
\newblock {\em The Astrophysical Journal} {\bf 2021}, {\em 912},~133.

\bibitem[Aloisio \em{et~al.}(2009)Aloisio, Berezinsky, and
  Gazizov]{aloisio2009}
Aloisio, R.; Berezinsky, V.; Gazizov, A.
\newblock The problem of superluminal diffusion of relativistic particles and
  its phenomenological solution.
\newblock {\em The Astrophysical Journal} {\bf 2009}, {\em 693},~1275.

\bibitem[Recchia \em{et~al.}(2021)Recchia, Di~Mauro, Aharonian, Orusa, Donato,
  Gabici, and Manconi]{recchia2021}
Recchia, S.; Di~Mauro, M.; Aharonian, F.A.; Orusa, L.; Donato, F.; Gabici, S.;
  Manconi, S.
\newblock Do the Geminga, Monogem and PSR J0622+ 3749 $\gamma$-ray halos imply
  slow diffusion around pulsars?
\newblock {\em Physical Review D} {\bf 2021}, {\em 104},~123017.

\bibitem[Cao \em{et~al.}(2021)Cao, Aharonian, An, Axikegu, Bai, Bai, Bao,
  Bastieri, Bi, et~al.]{crab2021}
Cao, Z.; Aharonian, F.; An, Q.; Axikegu.; Bai, L.; Bai, Y.; Bao, Y.; Bastieri,
  D.; Bi, X.;  et~al.
\newblock Peta--electron volt gamma-ray emission from the Crab Nebula.
\newblock {\em Science} {\bf 2021}, {\em 373},~425--430.

\bibitem[Fouka and Ouichaoui(2013)]{fouka2013}
Fouka, M.; Ouichaoui, S.
\newblock Analytical fits to the synchrotron functions.
\newblock {\em Research in Astronomy and Astrophysics} {\bf 2013}, {\em
  13},~680.

\bibitem[Prosekin \em{et~al.}(2015)Prosekin, Kelner, and
  Aharonian]{prosekin2015}
Prosekin, A.Y.; Kelner, S.R.; Aharonian, F.A.
\newblock Transition of propagation of relativistic particles from the
  ballistic to the diffusion regime.
\newblock {\em Physical Review D} {\bf 2015}, {\em 92},~083003.

\bibitem[{Schellenberger} \em{et~al.}(2015){Schellenberger}, {Reiprich},
  {Lovisari}, {Nevalainen}, and {David}]{Schellenberger2015}
{Schellenberger}, G.; {Reiprich}, T.H.; {Lovisari}, L.; {Nevalainen}, J.;
  {David}, L.
\newblock {XMM-Newton and Chandra cross-calibration using HIFLUGCS galaxy
  clusters . Systematic temperature differences and cosmological impact}.
\newblock {\em \aap} {\bf 2015}, {\em 575},~A30,
  \href{http://xxx.lanl.gov/abs/1404.7130}{{\normalfont
  [arXiv:astro-ph.CO/1404.7130]}}.
\newblock {\url{https://doi.org/10.1051/0004-6361/201424085}}.

\bibitem[Foreman-Mackey \em{et~al.}(2013)Foreman-Mackey, Hogg, Lang, and
  Goodman]{foreman2013}
Foreman-Mackey, D.; Hogg, D.W.; Lang, D.; Goodman, J.
\newblock emcee: the MCMC hammer.
\newblock {\em Publications of the Astronomical Society of the Pacific} {\bf
  2013}, {\em 125},~306.

\bibitem[{Atoyan} and {Aharonian}(1996)]{Atoyan1996}
{Atoyan}, A.M.; {Aharonian}, F.A.
\newblock {On the mechanisms of gamma radiation in the Crab Nebula}.
\newblock {\em \mnras} {\bf 1996}, {\em 278},~525--541.
\newblock {\url{https://doi.org/10.1093/mnras/278.2.525}}.

\bibitem[{Amato} \em{et~al.}(2003){Amato}, {Guetta}, and {Blasi}]{Amato2003}
{Amato}, E.; {Guetta}, D.; {Blasi}, P.
\newblock {Signatures of high energy protons in pulsar winds}.
\newblock {\em \aap} {\bf 2003}, {\em 402},~827--836,
  \href{http://xxx.lanl.gov/abs/astro-ph/0302121}{{\normalfont
  [arXiv:astro-ph/astro-ph/0302121]}}.
\newblock {\url{https://doi.org/10.1051/0004-6361:20030279}}.

\bibitem[{Zhang} and {Yang}(2009)]{Zhang09}
{Zhang}, L.; {Yang}, X.C.
\newblock {TeV Gamma-Ray Emission from Vela X: Leptonic or Hadronic?}
\newblock {\em \apjl} {\bf 2009}, {\em 699},~L153--L156.
\newblock {\url{https://doi.org/10.1088/0004-637X/699/2/L153}}.

\bibitem[{Li} \em{et~al.}(2010){Li}, {Chen}, and {Zhang}]{Li2010}
{Li}, H.; {Chen}, Y.; {Zhang}, L.
\newblock {Lepto-hadronic origin of {\ensuremath{\gamma}}-rays from the
  G54.1+0.3 pulsar wind nebula}.
\newblock {\em \mnras} {\bf 2010}, {\em 408},~L80--L84,
  \href{http://xxx.lanl.gov/abs/1008.2704}{{\normalfont
  [arXiv:astro-ph.HE/1008.2704]}}.
\newblock {\url{https://doi.org/10.1111/j.1745-3933.2010.00934.x}}.

\bibitem[{Di Palma} \em{et~al.}(2017){Di Palma}, {Guetta}, and
  {Amato}]{DiPalma2017}
{Di Palma}, I.; {Guetta}, D.; {Amato}, E.
\newblock {Revised Predictions of Neutrino Fluxes from Pulsar Wind Nebulae}.
\newblock {\em \apj} {\bf 2017}, {\em 836},~159,
  \href{http://xxx.lanl.gov/abs/1605.01205}{{\normalfont
  [arXiv:astro-ph.HE/1605.01205]}}.
\newblock {\url{https://doi.org/10.3847/1538-4357/836/2/159}}.

\bibitem[{Liu} and {Wang}(2021)]{Liu2021}
{Liu}, R.Y.; {Wang}, X.Y.
\newblock {PeV Emission of the Crab Nebula: Constraints on the Proton Content
  in Pulsar Wind and Implications}.
\newblock {\em \apj} {\bf 2021}, {\em 922},~221,
  \href{http://xxx.lanl.gov/abs/2109.14148}{{\normalfont
  [arXiv:astro-ph.HE/2109.14148]}}.
\newblock {\url{https://doi.org/10.3847/1538-4357/ac2ba0}}.

\bibitem[{Mori} \em{et~al.}(2022){Mori}, {An}, {Burgess}, {Capasso}, {Dingus},
  {Gelfand}, {Hailey}, {Humensky}, {Malone}, {Mukherjee}, {Park}, {Pope},
  {Reynolds}, {Safi-Harb}, {Woo}, and {Galactic TeV source
  Collaboration}]{Nustar2022}
{Mori}, K.; {An}, H.; {Burgess}, D.; {Capasso}, M.; {Dingus}, B.; {Gelfand},
  J.; {Hailey}, C.; {Humensky}, B.; {Malone}, K.; {Mukherjee}, R.;  et~al.
\newblock {NuSTAR broad-band X-ray observational campaign of energetic pulsar
  wind nebulae in synergy with VERITAS, HAWC and Fermi gamma-ray telescopes}.
\newblock In Proceedings of the 37th International Cosmic Ray Conference. 12-23
  July 2021. Berlin,  2022, p. 963,
  \href{http://xxx.lanl.gov/abs/2108.00557}{{\normalfont
  [arXiv:astro-ph.HE/2108.00557]}}.

\bibitem[Kafexhiu \em{et~al.}(2014)Kafexhiu, Aharonian, Taylor, and
  Vila]{kafexhiu2014}
Kafexhiu, E.; Aharonian, F.; Taylor, A.M.; Vila, G.S.
\newblock Parametrization of gamma-ray production cross sections for p p
  interactions in a broad proton energy range from the kinematic threshold to
  PeV energies.
\newblock {\em Physical Review D} {\bf 2014}, {\em 90},~123014.

\bibitem[{Tang} and {Chevalier}(2012)]{Tang2012}
{Tang}, X.; {Chevalier}, R.A.
\newblock {Particle Transport in Young Pulsar Wind Nebulae}.
\newblock {\em \apj} {\bf 2012}, {\em 752},~83,
  \href{http://xxx.lanl.gov/abs/1204.3913}{{\normalfont
  [arXiv:astro-ph.HE/1204.3913]}}.
\newblock {\url{https://doi.org/10.1088/0004-637X/752/2/83}}.

\bibitem[{Porth} \em{et~al.}(2016){Porth}, {Vorster}, {Lyutikov}, and
  {Engelbrecht}]{Porth2016}
{Porth}, O.; {Vorster}, M.J.; {Lyutikov}, M.; {Engelbrecht}, N.E.
\newblock {Diffusion in pulsar wind nebulae: an investigation using
  magnetohydrodynamic and particle transport models}.
\newblock {\em \mnras} {\bf 2016}, {\em 460},~4135--4149,
  \href{http://xxx.lanl.gov/abs/1604.03352}{{\normalfont
  [arXiv:astro-ph.HE/1604.03352]}}.
\newblock {\url{https://doi.org/10.1093/mnras/stw1152}}.

\bibitem[{Bao} \em{et~al.}(2019){Bao}, {Liu}, and {Chen}]{Bao2019}
{Bao}, Y.; {Liu}, S.; {Chen}, Y.
\newblock {On the Gamma-Ray Nebula of Vela Pulsar. I. Very Slow Diffusion of
  Energetic Electrons within the TeV Nebula}.
\newblock {\em \apj} {\bf 2019}, {\em 877},~54,
  \href{http://xxx.lanl.gov/abs/1907.02037}{{\normalfont
  [arXiv:astro-ph.HE/1907.02037]}}.
\newblock {\url{https://doi.org/10.3847/1538-4357/ab1908}}.

\bibitem[{Liu} and {Yan}(2020)]{Liu2020_1825}
{Liu}, R.Y.; {Yan}, H.
\newblock {On the unusually large spatial extent of the TeV nebula HESS
  J1825-137: implication from the energy-dependent morphology}.
\newblock {\em \mnras} {\bf 2020}, {\em 494},~2618--2627,
  \href{http://xxx.lanl.gov/abs/1907.02498}{{\normalfont
  [arXiv:astro-ph.HE/1907.02498]}}.
\newblock {\url{https://doi.org/10.1093/mnras/staa911}}.

\bibitem[{Hu} \em{et~al.}(2022){Hu}, {Ishizaki}, {Ng}, {Tanaka}, and
  {Mong}]{Hu2022}
{Hu}, C.P.; {Ishizaki}, W.; {Ng}, C.Y.; {Tanaka}, S.J.; {Mong}, Y.L.
\newblock {A Comprehensive Study of the Spectral Variation and the Brightness
  Profile of Young Pulsar Wind Nebulae}.
\newblock {\em \apj} {\bf 2022}, {\em 927},~87,
  \href{http://xxx.lanl.gov/abs/2201.09403}{{\normalfont
  [arXiv:astro-ph.HE/2201.09403]}}.
\newblock {\url{https://doi.org/10.3847/1538-4357/ac4d2d}}.

\bibitem[{Kennel} and {Coroniti}(1984)]{KC1984}
{Kennel}, C.F.; {Coroniti}, F.V.
\newblock {Magnetohydrodynamic model of Crab nebula radiation.}
\newblock {\em \apj} {\bf 1984}, {\em 283},~710--730.
\newblock {\url{https://doi.org/10.1086/162357}}.

\bibitem[{Van Etten} and {Romani}(2011)]{vanEtten2011}
{Van Etten}, A.; {Romani}, R.W.
\newblock {Multi-zone Modeling of the Pulsar Wind Nebula HESS J1825-137}.
\newblock {\em \apj} {\bf 2011}, {\em 742},~62,
  \href{http://xxx.lanl.gov/abs/1108.3573}{{\normalfont
  [arXiv:astro-ph.HE/1108.3573]}}.
\newblock {\url{https://doi.org/10.1088/0004-637X/742/2/62}}.

\bibitem[{Ishizaki} \em{et~al.}(2018){Ishizaki}, {Asano}, and
  {Kawaguchi}]{Ishizaki2018}
{Ishizaki}, W.; {Asano}, K.; {Kawaguchi}, K.
\newblock {Outflow and Emission Model of Pulsar Wind Nebulae with the Back
  Reaction of Particle Diffusion}.
\newblock {\em \apj} {\bf 2018}, {\em 867},~141,
  \href{http://xxx.lanl.gov/abs/1809.09054}{{\normalfont
  [arXiv:astro-ph.HE/1809.09054]}}.
\newblock {\url{https://doi.org/10.3847/1538-4357/aae389}}.

\bibitem[{Hillas}(1984)]{Hillas1984}
{Hillas}, A.M.
\newblock {The Origin of Ultra-High-Energy Cosmic Rays}.
\newblock {\em \araa} {\bf 1984}, {\em 22},~425--444.
\newblock {\url{https://doi.org/10.1146/annurev.aa.22.090184.002233}}.

\bibitem[Gaensler and Slane(2006)]{gaensler2006}
Gaensler, B.M.; Slane, P.O.
\newblock The evolution and structure of pulsar wind nebulae.
\newblock {\em Annu. Rev. Astron. Astrophys.} {\bf 2006}, {\em 44},~17--47.

\bibitem[{De Angelis} \em{et~al.}(2017){De Angelis}, {Tatischeff}, {Tavani},
  {Oberlack}, {Grenier}, {Hanlon}, {Walter}, {Argan}, {von Ballmoos},
  {Bulgarelli}, {Donnarumma}, {Hernanz}, {Kuvvetli}, {Pearce}, {Zdziarski},
  {Aboudan}, {Ajello}, {Ambrosi}, {Bernard}, {Bernardini}, {Bonvicini},
  {Brogna}, {Branchesi}, {Budtz-Jorgensen}, {Bykov}, {Campana}, {Cardillo},
  {Coppi}, {De Martino}, {Diehl}, {Doro}, {Fioretti}, {Funk}, {Ghisellini},
  {Grove}, {Hamadache}, {Hartmann}, {Hayashida}, {Isern}, {Kanbach}, {Kiener},
  {Kn{\"o}dlseder}, {Labanti}, {Laurent}, {Limousin}, {Longo}, {Mannheim},
  {Marisaldi}, {Martinez}, {Mazziotta}, {McEnery}, {Mereghetti}, {Minervini},
  {Moiseev}, {Morselli}, {Nakazawa}, {Orleanski}, {Paredes}, {Patricelli},
  {Peyr{\'e}}, {Piano}, {Pohl}, {Ramarijaona}, {Rando}, {Reichardt},
  {Roncadelli}, {Silva}, {Tavecchio}, {Thompson}, {Turolla}, {Ulyanov},
  {Vacchi}, {Wu}, and {Zoglauer}]{eASTROGAM2017}
{De Angelis}, A.; {Tatischeff}, V.; {Tavani}, M.; {Oberlack}, U.; {Grenier},
  I.; {Hanlon}, L.; {Walter}, R.; {Argan}, A.; {von Ballmoos}, P.;
  {Bulgarelli}, A.;  et~al.
\newblock {The e-ASTROGAM mission. Exploring the extreme Universe with gamma
  rays in the MeV - GeV range}.
\newblock {\em Experimental Astronomy} {\bf 2017}, {\em 44},~25--82,
  \href{http://xxx.lanl.gov/abs/1611.02232}{{\normalfont
  [arXiv:astro-ph.HE/1611.02232]}}.
\newblock {\url{https://doi.org/10.1007/s10686-017-9533-6}}.

\bibitem[{Moiseev} and {Amego Team}(2017)]{AMEGO2017}
{Moiseev}, A.; {Amego Team}.
\newblock {All-Sky Medium Energy Gamma-ray Observatory (AMEGO)}.
\newblock In Proceedings of the 35th International Cosmic Ray Conference
  (ICRC2017),  2017, Vol. 301, {\em International Cosmic Ray Conference}, p.
  798.

\end{thebibliography}

%


\end{adjustwidth}
\end{document}